\documentclass[12pt]{article}
\usepackage{epsf}
\usepackage{cite}
\usepackage{amssymb}
\newcommand{\be}{\begin{equation}}
\newcommand{\ee}{\end{equation}}
\newcommand{\bea}{\begin{eqnarray}}
\newcommand{\eea}{\end{eqnarray}}
\newcommand{\ba}{\begin{array}}
\newcommand{\ea}{\end{array}}

\newcommand{\pl}{Phys.\ Lett.}
\newcommand{\np}{Nucl.\ Phys.}
\newcommand{\prl}{Phys.\ Rev.\ Lett.}
\newcommand{\prd}{Phys.\ Rev.\ D}
\newcommand{\pr}{Phys.\ Rev.}

\newcommand{\mpl}{Mod.\ Phys.\ Lett.} 
\newcommand{\etal}{{\it et al}.,\ }

\begin{document}
\title{The $B\rightarrow X_s\gamma\gamma$ rare decay}
\author{\renewcommand{\thefootnote}{\alph{footnote}}
L. Reina\footnotemark[1], G. Ricciardi\footnotemark[2], 
A. Soni\footnotemark[1]\\
\renewcommand{\thefootnote}{\alph{footnote}}
\footnotemark[1] Physics Department, Brookhaven National Laboratory,\\
 Upton, NY 11973 \\
\renewcommand{\thefootnote}{\alph{footnote}}
\footnotemark[2] Dipartimento di Scienze Fisiche, Universit\`a degli 
Studi di Napoli,\\
and I.N.F.N., Sezione di Napoli,\\
Mostra d'Oltremare, Pad. 19, I-80125 Napoli, Italy }
\date{}
\maketitle
\renewcommand{\thefootnote}{\arabic{footnote}}
\begin{abstract}
The rare decay $B\rightarrow X_s\gamma\gamma$ is studied in the
Standard Model (SM) and in two different versions (Model I and Model
II) of the Two Higgs Doublet Model (2HDM). In the SM the branching
ratio into \emph{hard photons} is about $1\times 10^{-7}$ and can be
appreciably different in the 2HDM. We also introduce a
forward-backward asymmetry which gives an additional handle to
discriminate different models.
\end{abstract} 

\baselineskip 20pt
\section{Introduction}

Intense experimental effort is being directed to B-physics. Several
new facilities are on the horizon. There are, of course, the $e^+e^-$
based symmetric and asymmetric B-factories at Cornell, KEK and SLAC,
in addition to LEP II. Progress is also being made in the hadronic
environment at HERA-B and there are some plans for TEV-B and far into
the future, LHC-B. Indeed it is useful to recall that in the case of
the kaon some of the branching ratios currently being measured have
reached the $10^{-10}$ level. It should then be clear that in
B-physics too, the experimental activity is likely to continue to
flourish well beyond the presently attainable branching ratios of
about $10^{-5}$.

Amongst the rare decays the flavor changing decays of the B meson are
of special interest, in particular those driven by the electroweak
(EW) penguins, due to their relative cleanliness and to their
sensitivity to new physics \cite{hewett}. In this category, the two
decays that have dominated the discussion for over a decade are
$b\rightarrow s\gamma$ and $b\rightarrow s l^+l^-$. It is useful to
recall that the inclusive branching ratio for $b\rightarrow s\gamma$
has been measured at CLEO to be \cite{alam}

\be
Br(B\rightarrow X_s\gamma)^\mathrm{exp}=(2.32\pm 0.51\pm 0.29\pm 0.32)
\times 10^{-4}
\label{bsg_exp}
\ee

\noindent and a Next-to-Leading Order calculation of the same now
exists \cite{greub,misiak,yao1,bsgall}. The result \cite{misiak}:
$Br(B\rightarrow X_s\gamma)^\mathrm{th}= (3.28\pm 0.33)\times 10^{-4}$
indicates agreement of the SM prediction with the CLEO measurement
within $2\sigma$.  As far as $b\rightarrow sl^+l^-$ is concerned, the
non-resonant part of $B\rightarrow X_s\mu^+\mu^-$ has been predicted
to give \cite{munz}: $Br(B\rightarrow X_s\mu^+\mu^-)_{NR}= (5.7\pm
0.9)\times 10^{-6}$, while the existing upper bound on this decay mode
from D$\emptyset$ is now being updated to \cite{buras}

\be
Br(B\rightarrow X_s\mu^+\mu^-)\le 3.2\times 10^{-5}\,\,\,\,,
\label{bsmumu_exp}
\ee

\noindent less than one order of magnitude away from the theoretical
prediction. 

Our focus in this paper will be on a related mode: $b\rightarrow
s\gamma\gamma$ which is of the same order in the EW couplings as
$b\rightarrow s l^+l^-$ and is also of considerable interest. We will
extend earlier works on this process and study it in the SM and in one
of its most popular extensions, namely the 2HDM. The calculation of
$b\rightarrow s\gamma\gamma$ in the SM at the lowest order in the EW
interactions and without QCD corrections gives a branching ratio of
about $10^{-7}$, therefore in the ballpark of the rare B decays that
should become accessible at the future B-meson
facilities. Furthermore, we will introduce a forward-backward
asymmetry which should be less sensitive to QCD corrections than the
branching ratio and should also constitute a useful probe of the
theory.

\section{The $\lowercase{b}\rightarrow\lowercase{s\gamma\gamma}$
decay in the Standard Model}

\begin{figure}[ht]
\centering
\epsfxsize=5.in
\leavevmode\epsffile{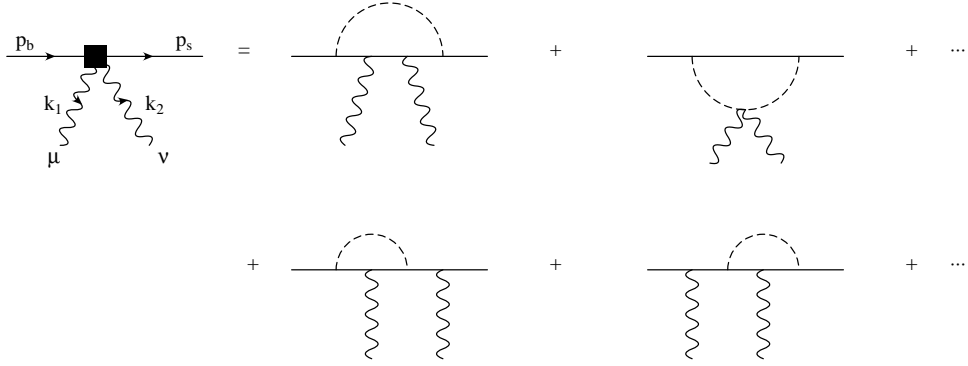}
\caption[]{Examples of 1PI and 1PR contributions to the $b\rightarrow
s\gamma\gamma$ process. The dashed line represents a \emph{W} (SM) or a
charged scalar (2HDM).}
\label{bsgg}
\end{figure}

The decay $B\rightarrow X_s\gamma\gamma$\footnote{The inclusive
calculation presented here is applicable to $B_u$, $B_d$ and $B_s$
with the corresponding multiparticle state ($X_s$) having the flavors
of $\bar s u$, $\bar s d$ and $\bar s s$ respectively. Moreover our
calculations can be readily adapted to the case of $b\rightarrow
d\gamma\gamma$ with obvious changes.} can be studied to a very good
approximation in terms of the quark level decay $b\rightarrow
s\gamma\gamma$ \cite{bigi}. The total amplitude for the quark level
process can be written as \cite{yao2,simma,herrlich}

\be
A(b\rightarrow s\gamma\gamma)=-i\frac{\alpha_e G_F}{\sqrt{2}\pi}
\epsilon_{\mu}(k_1)\epsilon_{\nu}(k_2)\bar u(p_s)T^{\mu\nu}u(p_b)
\,\,\,\,,
\label{ampl_bbgg}
\ee

\noindent where $\alpha_e$ is the electromagnetic fine structure
constant, $G_F$ the Fermi coupling constant, $p_b$ and $p_s$ denote
the momenta of the incoming and outgoing quarks and
$\epsilon_{\mu}(k_1)$ and $\epsilon_{\nu}(k_2)$ are the polarization
vectors of the two photons.  The tensor $T^{\mu\nu}$ is derived from
the calculation of the Feynman diagrams of Fig.~\ref{bsgg}, when we
sum over the three possible flavors of quarks that run in the penguin
loop

\be
T^{\mu\nu}=\sum_{i=u,c,t}\lambda_i
T_i^{\mu\nu}=\lambda_u(T^{\mu\nu}_u-T^{\mu\nu}_c)
+\lambda_t(T^{\mu\nu}_t-T^{\mu\nu}_c) \,\,\,\,.
\label{t_munu}
\ee

\noindent Here $\lambda_i=V_{ib}V^*_{is}$ is the CKM factor
multiplying the loop diagrams with an internal $i$-quark and we have
used the unitarity of the CKM matrix to obtain the last result. Even
though $\lambda_u\ll\lambda_t$, for our purpose, as we will explain
later, the first term cannot be neglected.

As we see from Fig.~1, both one-particle reducible (1PR) and
one-particle irreducible (1PI) diagrams contribute to the process. We
will follow the notation of Ref.~\cite{herrlich} and write the total
amplitude for the \emph{i} flavor as

\be
T^{\mu\nu}_i=T^{\mu\nu}_{i,\mathrm{1PI}}+T^{\mu\nu}_{i,\mathrm{1PR}}
\,\,\,.
\label{t_munu_i}
\ee

\noindent The 1PI  contribution $T^{\mu\nu}_{i,\mathrm{1PI}}$ is then
written as

\be
T^{\mu\nu}_{i,\mathrm{1PI}}=\frac{8}{9}\delta_3(z_i)I^{\mu\nu}
\,\,\,\,,
\ee

\noindent where $z_i=2k_1\cdot k_2/m_i^2$ and $\delta_3(z_i)$ is a
function defined by \cite{gaillard} 

\be
\delta_3(z_i)=1+\frac{2}{z_i}\int_0^1\frac{du}{u}
\log\left[1-z_iu(1-u)\right]\,\,\,.
\label{delta3}
\ee

\noindent The tensor $I^{\mu\nu}$ is given by\footnote{We want to
emphasize here that both Refs.~\cite{yao2} and \cite{herrlich} deal
with the decay $B_s\rightarrow\gamma\gamma$. In that case one only
needs to keep the first term in the $I^{\mu\nu}$ tensor. This is not
true for the inclusive $B\rightarrow X_s\gamma\gamma$ decay.}

\be
I^{\mu\nu}=\left[
i\epsilon^{\mu\nu\xi\alpha}\gamma_{\alpha}\gamma_L (k_1-k_2)_{\xi}+
i\frac{k_{1\xi}k_{2\eta}}{k_1\cdot k_2}
(\epsilon^{\mu\xi\eta\alpha}k_1^{\nu}-
\epsilon^{\nu\xi\eta\alpha}k_2^{\mu})\gamma_{\alpha}\gamma_L\right]
\,\,\,\,,
\label{t_munu_1pi}
\ee

\noindent where we use the notation $\gamma_{L,R}=(1\mp\gamma_5)/2$.
On the other hand, the 1PR contribution has the form

\be 
T^{\mu\nu}_{i,\mathrm{1PR}}=-i\frac{1}{3}F_2(x_i)W^{\mu\nu}
(m_b\gamma_R+m_s\gamma_L)\,\,\,\,,
\ee

\noindent where $F_2(x_i)$ for $x_i=m_i^2/M_W^2$ is the 
form factor of the magnetic dipole operator of the $bs\gamma$ vertex
\cite{inamilim}, given by

\be
F_2(x_i)=\frac{1}{36(1-x_i)^4}\left[-46+205x_i-312x_i^2+175x_i^3-
22x_i^4+(36x_i^2-54x_i^3)\log(x_i))\right]\,\,\,,
\label{f2}
\ee

\noindent while the $W^{\mu\nu}$ tensor is defined by the expression

\bea
\label{t_munu_1pr}
W^{\mu\nu}&=&-\left[\left(\frac{p_s^{\nu}}{p_s\cdot
k_2}-\frac{p_b^{\nu}}{p_b\cdot
k_2}\right)\sigma(\mu,k_1)+\left(\frac{p_s^{\mu}}{p_s\cdot
k_1}-\frac{p_b^{\mu}}{p_b\cdot k_1}\right)\sigma(\nu,k_2)\right]\\
&+&\frac{i}{2}\left[\left(\frac{1}{p_s\cdot k_2}-\frac{1}{p_b\cdot
k_1}\right)\sigma(\nu,k_2)\sigma(\mu,k_1)+\left(\frac{1}{p_s\cdot
k_1}-\frac{1}{p_b\cdot
k_2}\right)\sigma(\mu,k_1)\sigma(\nu,k_2)\right]\,\,\,.\nonumber 
\eea

The rate $\Gamma(b\rightarrow s\gamma\gamma)$ can therefore be
decomposed as the sum of a pure 1PR, a pure 1PI and an interference
1PR-1PI term, i.e.

\be
\Gamma(b\rightarrow s\gamma\gamma)=\Gamma_\mathrm{1PR}+
\Gamma_\mathrm{1PI}+\Gamma_\mathrm{1PR-1PI}\,\,\,\,.
\label{gamma}
\ee

\noindent In order to obtain the total rate into \emph{hard photons}
we have to place suitable kinematical cuts and perform one dimensional
integration numerically. We have also checked our results integrating
over the phase space with a Montecarlo event generator.

Indeed the total rate and the relevance of each different contribution
(1PR,1PI) depends appreciably on the kinematical cuts imposed. In
order to isolate two \emph{hard} photons, we demand that their energy
is not too small and that they are not too collinear to each other and
to the outgoing \emph{s} quark. We thus require the energy of each
photon to be larger than 100 MeV and that $s_1=(p_s+k_1)^2$,
$s_2=(k_1+k_2)^2$ and $s_3=(p_s+k_2)^2$ satisfy

\be
s_1\ge c\,m_b^2\,\,\,\,\,,\,\,\,\,\, s_2\ge c\,m_b^2\,\,\,\,\,,
\,\,\,\,\,s_3\ge c\,m_b^2\,\,\,.
\label{cuts}
\ee

\noindent We take $c=0.01$ and $c=0.02$ to study the dependence on the
cuts. Note that the resulting two photon invariant mass squared is at
least one order of magnitude bigger than $m_\pi^2$. Furthermore all
the angles are taken to be $\gtrsim 20^\circ$.

\begin{table}[ht]
\begin{center}
\begin{tabular}{|l||c|c c c c|c|}
\hline\hline
& &\multicolumn{4}{|c|}{\rule[-3mm]{0mm}{8mm} 
Br$(b\rightarrow s\gamma\gamma)\times 10^{-7}$} &  \\ 
& & & 1-Particle & 1-Particle & & $A_\mathrm{FB}$\\ 
& & \raisebox{1.5ex}[0pt]{Total} & Reducible & Irreducible &
\raisebox{1.5ex}[0pt]{Interference} & \\ \hline
& & \multicolumn{5}{|c|}{\rule[-2mm]{0mm}{6mm}c=0.01}\\ \cline{3-7}
{\rule[-2mm]{0mm}{8mm}SM} & & 1.60 & 1.30 & 0.23 & 0.08 & 0.69 \\ 
\cline{2-2}
&{\rule[-2mm]{0mm}{6mm}$\tan\beta$} & & & & & \\ \cline{2-2}
{\rule[-2mm]{0mm}{8mm}2HDM} & 0.5 & 0.23-14.67 & 0.003-14.70 & 0.23 &
-(0.26-0.003) & 0.37-0.77 \\
Model I & 1 & 0.23-1.26 & 0.01-0.96 & 0.23 & -0.007-0.07 & 0.39-0.67 \\
& 10 & 1.57-1.59 & 1.26-1.29 & 0.23 & 0.07 & 0.68 \\ 
{\rule[-2mm]{0mm}{8mm}2HDM} & 0.5 & 2.19-16.67 & 1.87-16.17 & 0.23 & 
0.09-0.27 & 0.70-0.75 \\
Model II & 1 & 2.07-9.65 & 1.75-9.21 & 0.23 & 0.09-0.21 & 0.70-0.74 \\
& 10 & 2.03-7.76 & 1.71-7.34 & 0.23 & 0.09-0.18 & 0.70-0.74 \\ \hline
& & \multicolumn{5}{|c|}{\rule[-2mm]{0mm}{6mm}c=0.02}\\ \cline{3-7}
{\rule[-2mm]{0mm}{8mm}SM} & & 1.33 & 1.02 & 0.23 & 0.08 & 0.64\\ 
\cline{2-2}
&{\rule[-2mm]{0mm}{6mm}$\tan\beta$} & & & & & \\ \cline{2-2}
{\rule[-2mm]{0mm}{8mm}2HDM} & 0.5 & 0.23-11.57 & 0.003-11.60 & 0.23 &
-(0.26-0.003)& 0.37-0.73\\
Model I & 1 & 0.23-1.05 & 0.009-0.76 & 0.23 & -0.007-0.07 & 0.38-0.62 \\
& 10 & 1.30-1.33 & 0.99-1.01 & 0.23 & 0.07 & 0.64\\ 
{\rule[-2mm]{0mm}{8mm}2HDM} & 0.5 & 1.80-13.26 & 1.48-12.76 & 0.23 &
0.09-0.27 & 0.66-0.71\\
Model II & 1 & 1.70-7.71 & 1.38-7.23 & 0.23 & 0.09-0.21 & 0.65-0.71 \\
& 10 & 1.67-6.21 & 1.35-5.79 & 0.23 & 0.089-0.18 & 0.65-0.70\\ 
\hline\hline
\end{tabular}
\caption[]{Values of branching ratio and forward-backward asymmetry
obtained in the SM and in the 2HDM's (Model I and II), for two
different values of the cut. In the 2HDM case we give a range of
values corresponding to the variation of $M_h$ between 100 GeV and 1
TeV.}
\label{brall}
\end{center}
\end{table}

Our results are summarized in Table~\ref{brall}.  These results are
for pure EW penguins and do not include QCD corrections
\cite{bsgg2}. For $m_t\simeq 175$ GeV, as in the $b\rightarrow
s\gamma$ case, QCD corrections are not likely to change the
predictions of Table~\ref{brall} dramatically. The branching ratios
are calculated by normalizing to the semileptonic branching ratio, for
which we have used the experimental value $Br(B\rightarrow X_s
e\nu_e)\simeq 0.11$. We did not include QCD corrections in the
semileptonic rate either; for our purpose QCD effects are completely
marginal here. For the numerical results we have used: $m_t=175$ GeV,
$m_c=1.5$ GeV, $m_u=4.5$ MeV and $m_b=4.5$ GeV. However, we have
checked the dependence of the results on $m_c$ and $m_u$, allowing
them to vary in the ranges $1.2<m_c<1.8$ GeV and $4.5<m_u<100$ MeV. We
find that the dependence on $m_u$ is totally negligible while the
branching ratio varies up to $15-20\%$ within this very conservative
range of $m_c$. For $m_s$ we have made the following distinction: in
the calculation of the quark decay amplitude we have taken the current
quark mass, $m_s=150$ MeV, while in the integration over the phase
space, in order to respect the kinematics of the decay as closely as
we can, we have used the $K$ meson mass and taken approximately
$m_s=450$ MeV. Again this assumption does not affect the results by
more than $8\%$ in any case. As far as the CKM matrix elements are
concerned, we have used: $V_{cs}\simeq V_{tb}\simeq 1$, $V_{us}\simeq
0.22$, $|V_{ts}|\simeq |V_{cb}|\simeq 0.04$ and $|V_{ub}/V_{cb}|\simeq
0.08$.

In Table~\ref{brall} we also give results separately due to 1PR, 1PI
and due to their interference. As we can see, the role played by the
1PR and 1PI contributions varies a little with the different cuts
imposed.  With mild cuts, the 1PR contribution is dominant, due to the
divergent behavior of the amplitude for $s_2\rightarrow 0$. This
divergent behavior is caused by the presence of both infrared
($E_{\gamma}\rightarrow 0$) and collinear
($\cos\theta_{\gamma\gamma}\rightarrow 0$) divergencies, which, for
sure, would cancel in a complete analytical calculation when virtual
photon corrections at the same order in $\alpha_e$ are also
included. Experimentally, if one of the two photons in $b\rightarrow
s\gamma\gamma$ decay is emitted with very low energy and/or at a very
small angle with respect to the other photon, it becomes difficult to
distinguish the event from a $b\rightarrow s\gamma$ decay. Therefore
we exclude this region of the phase space by imposing the cuts
mentioned earlier. This tends to suppress the 1PR contribution
compared to the 1PI one.

\begin{figure}[ht]
\centering
\epsfxsize=4.in
\leavevmode\epsffile{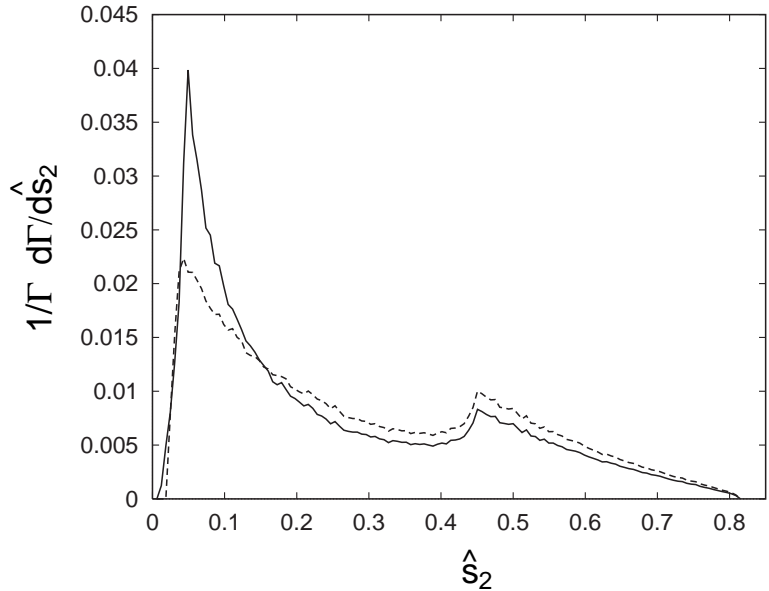}
\caption[]{The normalized distribution $1/\Gamma\cdot
d\Gamma/d\hat{s}_2$ versus $\hat{s}_2=s_2/m_b^2$ in the SM for two
values of the cut parameter: $c=0.01$ (solid) and $c=0.02$
(dashed). Also, $\langle\hat{s}_2\rangle= 0.25,0.28$ for $c=0.01$ or
$c=0.02$ respectively.}
\label{dgammads2}
\end{figure}

Moreover, the analytical expressions of the 1PR and 1PI terms in
Eqs.~(\ref{t_munu_i})-(\ref{t_munu_1pr}) are such that the light
quarks mainly contribute to the 1PI part and the heavy quarks to the
1PR one. This is due to the analytical behavior of the coefficient
functions $\delta_3(z_i)$ and $F_2(x_i)$ (see Ref.~\cite{herrlich} for
a plot of these functions). The physical reason for this is that, when
the top quark runs in the loop of the penguin diagram for
$b\rightarrow s\gamma\gamma$ (both in the 1PR and in the 1PI
diagrams), the amplitude for the decay is very well approximated by
the amplitude for the emission of a brehmstrahlung photon off the
\emph{s} quark leg. Indeed by this reasoning, the authors of
Ref.~\cite{yao1} were able to get the right expression for the top
contribution using Low's theorem \cite{low}. Thus the higher the
kinematical cuts imposed to reduce the \emph{s} quark brehmstrahlung
events, the higher the relevance of the light quark contribution
(i.e. 1PI) to the total amplitude. For this reason we retained in
Eq.~(\ref{t_munu}) both the $\lambda_t$ and the $\lambda_u$ terms, in
order to have the light quark dependence fully under control.

\begin{figure}[ht]
\centering
\epsfxsize=4.in
\leavevmode\epsffile{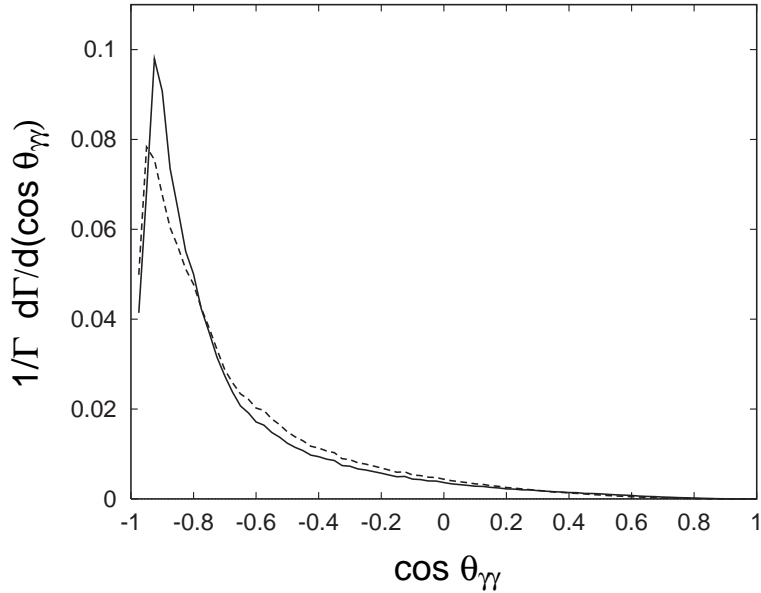}
\caption[]{The normalized distribution $1/\Gamma\cdot
d\Gamma/d(\cos\theta_{\gamma\gamma})$ versus
$\cos\theta_{\gamma\gamma}$ in the SM for two values of the cut
parameter: $c=0.01$ (solid) and $c=0.02$ (dashed). Also,
$\langle\cos\theta_{\gamma\gamma}\rangle= -0.70,-0.68$ for $c=0.01$ or
$c=0.02$ respectively.}
\label{dgammadcosth}
\end{figure}

In Fig.~\ref{dgammads2} and \ref{dgammadcosth} we present the results
for two different distributions: $d\Gamma/d\hat{s}_2$
($\hat{s}_2=s_2/m_b^2$) and $d\Gamma/d(\cos\theta_{\gamma\gamma})$,
where $\theta_{\gamma\gamma}$ is the angle between the two photons,
for the two sets of cuts\footnote{We observe that the behavior of
$1/\Gamma\cdot d\Gamma/d\hat{s}_2$ around $\hat{s}_2\simeq 0.45$
(i.e. $s_2\simeq 9\,\mbox{GeV}^2$) corresponds, for $m_c=1.5$ GeV, to
$s_2=4m_c^2$, i.e. to the physical threshold for an on-shell $c\bar c$
pair.}. As we can see the kinematics of the process is such that the
{\emph s} quark tends to be emitted with rather high energy,
compensated by the harder of the two photons, while the less energetic
photon tends to go in the direction of the {\emph s quark}. This
topology is typical of a brehmstrahlung event of the \emph{s} quark
and in fact it gets enhanced when we choose milder cuts on the
energies and on the angles.

\vspace{.5cm} Although the $b\rightarrow s\gamma\gamma$ process is
higher order in $\alpha_e$ and rarer than $b\rightarrow s\gamma$, it
does allow one to introduce an additional feature that may be a useful
probe of the dynamics of the decay, namely a forward-backward
asymmetry, defined as

\be
A_\mathrm{FB}=\frac{\Gamma(\cos\theta_{s\gamma}\ge
0)-\Gamma(\cos\theta_{s\gamma}<0)} {\Gamma(\cos\theta_{s\gamma}\ge
0)+\Gamma(\cos\theta_{s\gamma}<0)}\,\,\,\,,
\label{asym}
\ee

\noindent where $\theta_{s\gamma}$ is the angle between the \emph{s}
quark and the softer photon. Indeed, due to the identical nature of
the two photons, this is the only non-trivial angle we can think of to
study a forward-backward asymmetry. We recall that a forward-backward
asymmetry has also been found to be useful in the study of the
$b\rightarrow s l^+l^-$ decay \cite{ali}. Our results are summarized
in Table~\ref{brall}, for two different values of the cuts.  In
passing, we must remark that we expect $A_\mathrm{FB}$ to be less
affected by QCD corrections than the branching ratio.

\section{The $\lowercase{b}\rightarrow\lowercase{s\gamma\gamma}$
decay in the Two Higgs Doublet Model}

We want now to consider the $b\rightarrow s\gamma\gamma$ decay in the
context of a 2HDM with no flavor changing neutral currents allowed at
the tree level, i.e. Model I and Model II, in which the up and down
type quarks couple respectively to the same or to two different Higgs
doublets.

\begin{figure}[ht]
\centering
\epsfxsize=4.in
\leavevmode\epsffile{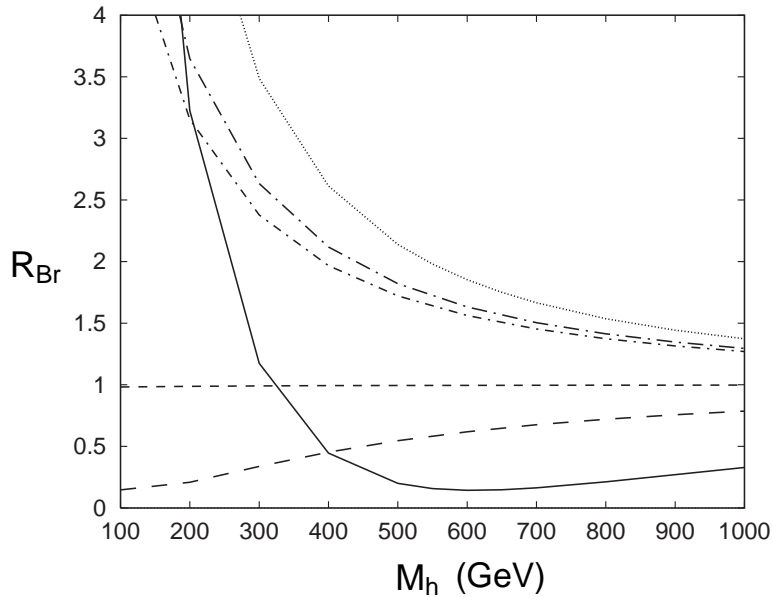}
\caption[]{The ratio $R_\mathrm{Br}$ as a function of the charged
scalar mass $M_h$, for different values of $\tan\beta$ in Model I
(solid: $\tan\beta=0.5$; long-dashed: $\tan\beta=1$; short-dashed:
$\tan\beta=10$) and Model II (dotted: $\tan\beta=0.5$;
long-dash-dot: $\tan\beta=1$; short-dash-dot: $\tan\beta=10$). The
curves correspond to the cut parameter $c=0.01$.}
\label{Rbr}
\end{figure}

The physical charged scalar field contributes to the $b\rightarrow
s\gamma\gamma$ decay via a new class of Feynman diagrams in which the
$W$ boson is replaced by the charged scalar. However, the only
important contributions are those with a top quark in the penguin
loop, because of the proportionality of the scalar-fermion coupling to
the fermion mass\footnote{This means that the new contribution to the
$m_c$ and $m_u$-dependent terms in the amplitude are much smaller than
the corresponding SM contributions, both 1PR and 1PI.}. Therefore, the
amplitude for $b\rightarrow s\gamma\gamma$ in Model I and Model II
will be given by Eq.~(\ref{t_munu}) where we only modify
$T^{\mu\nu}_{t,\mathrm{1PR}}$ to include the new form factor
\cite{aliev}

\bea
F_2^\mathrm{2HDM}(y_t)&=&\frac{y_t}{36(1-y_t)^4}\left\{
\xi^2\left[7-12y_t-3y_t^2+8y_t^3+6y_t(2-3y_t)\log y_t\right]+\right.
\\ \nonumber
&&6\left.\xi\xi^\prime(1-y_t)\left[3-8y_t+5y_t^2+2(2-3y_t)\log
y_t\right]\right\}\,\,\,\,,
\label{f2_2hdm}
\eea

\noindent where $y_t=(m_t/M_h)^2$ and $M_h$ denotes the mass of the
charged scalar. Moreover we have used the notation\footnote{Note that
the notation of Ref.~\cite{aliev} for Model I and Model II is the
reverse of ours.}: $\xi=v_1/v_2=1/\tan\beta$ and respectively

\bea
\label{modelIandII}
\xi^\prime&=& \xi\,\,\,\,\,\,\,\,\mbox{in Model I}\\
\xi^\prime&=&-\frac{1}{\xi}\,\,\,\,\,\mbox{in Model II}\,\,\,\,.\nonumber
\eea

\begin{figure}[ht]
\centering
\epsfxsize=4.in
\leavevmode\epsffile{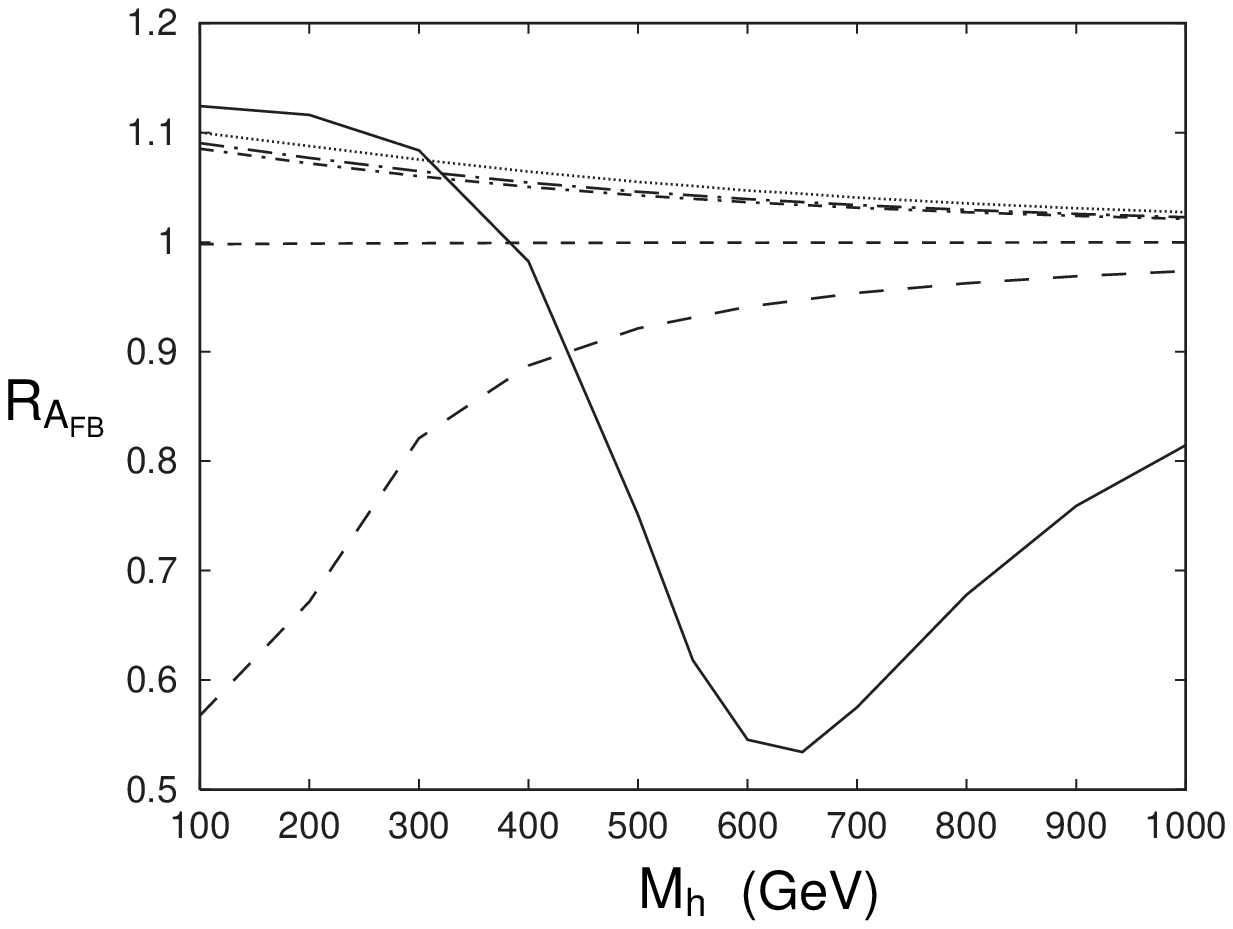}
\caption[]{The ratio $R_\mathrm{A_\mathrm{FB}}$ as a function of the
charged scalar mass $M_h$, for different values of $\tan\beta$ in
Model I (solid: $\tan\beta=0.5$; long-dashed: $\tan\beta=1$;
short-dashed: $\tan\beta=10$) and Model II (dotted: $\tan\beta=0.5$;
long-dash-dot: $\tan\beta=1$; short-dash-dot: $\tan\beta=10$). The
curves correspond to the cut parameter $c=0.01$.}
\label{Rasym}
\end{figure}
\noindent Summary of comparisons is given in Table~\ref{brall} and in
Figs.~\ref{Rbr}-\ref{Rasym}. In these two figures the differences
between the 2HDM's and the SM are parametrized in terms of the two
ratios

\be 
R_\mathrm{Br}=\frac{Br(B\rightarrow X_s\gamma\gamma)^\mathrm{2HDM}}
{Br(B\rightarrow X_s\gamma\gamma)^\mathrm{SM}}\,\,\,\,\,\,\,\,\,
\mbox{and}\,\,\,\,\,\,\,\,
R_\mathrm{A_\mathrm{FB}}=\frac{A_\mathrm{FB}^\mathrm{2HDM}}
{A_\mathrm{FB}^\mathrm{SM}}
\ee

\noindent both for Model I and Model II. For illustrative purposes we
consider three values of $\tan\beta$, namely $\tan\beta=0.5,1,10$ and
we allow $M_h$ to range between 100 GeV and 1 TeV. In this regard we
should recall that $M_h$ is also somewhat constrained by the
experimental measurement of $Br(B\rightarrow X_s\gamma)$ (see
Eq.~(\ref{bsg_exp})). The present situation seems to indicate that
$M_h\gtrsim 200-300$ GeV. However both experiment and theory still
have an appreciable error. Therefore we are tentatively considering
$M_h\ge 100$ GeV. The upper bound of 1 TeV is, of course, dictated by
the requirement of a weak-coupled scalar sector.  For Figs.~\ref{Rbr}
and \ref{Rasym} we also fixed $c=0.01$ and kept the other cuts on
$E_\gamma$ and on the angles as before. In fact a different value of
the cut parameter (e.g. $c=0.02$) does not change the curves in
Figs.~\ref{Rbr} and \ref{Rasym} significantly.

We can see that Model II gives always both a bigger branching ratio
and a bigger asymmetry than the SM. However while the asymmetry is
very close to the SM value, the branching ratio can be much larger
than the SM one over a wide range of values of the scalar mass,
especially for light $M_h$.

The behavior of Model I is quite different. Both the branching
ratio\footnote{A similar behavior was found in Ref.~\cite{hou} for the
$b\rightarrow s\gamma$ decay.}  and the asymmetry may be bigger or
smaller than the SM, over different ranges of $M_h$ and for different
values of $\tan\beta$. Moreover the relative importance of 1PR and 1PI
contributions may be very different than in the SM, due to a
cancellation between $F_2^\mathrm{SM}(x_t)$ and $F_2^\mathrm{
2HDM}(y_t)$ for some particular values of $M_h$ and $\tan\beta$. As we
can see from Table~\ref{brall}, there are cases in which the 1PI
contribution (i.e. mainly the light quark one) is dominant.  It is
also interesting to note how different values of $\tan\beta$ imply a
totally different behavior both of $R_\mathrm{Br}$ and of
$R_\mathrm{A_\mathrm{FB}}$ and they complement each other by giving
substantial deviations from the SM in different regions of the $M_h$
spectrum. For example, in the case of $\tan\beta=0.5$, for light $M_h$
the branching ratio gives a much larger deviation from the SM than the
asymmetry, while for heavier $M_h$, even up to $M_h\simeq 1$ TeV, the
asymmetry can be more interesting, especially because the branching
ratio is smaller than the SM one in this region. For $\tan\beta=1$, on
the other hand, both the branching ratio and the asymmetry are quite
different from the SM for $M_h\lesssim 500$ GeV.

This description has one obvious shortcoming, namely that QCD
corrections are not included. However we do not think that the
inclusion of QCD corrections in either the SM or in the 2HDM
calculation should greatly change the important features that we have
decribed above.

\section{Experimental Considerations}

Our primary focus of course has been on the inclusive $b\rightarrow
s\gamma\gamma$ process. As always it is very difficult to make
reliable predictions about exclusive channels. Nevertheless, some
general remarks can be made. Unlike $b\rightarrow s\gamma$, for
$b\rightarrow s\gamma\gamma$ the pseudoscalar ($K$ for $B$, $\eta$ and
$\eta^\prime$ for $B_s$) as well as the vector ($K^\ast$ for $B$ and
$\phi$ for $B_s$) final states are both allowed. These single meson
states are likely to be a large fraction, perhaps several tens of
percents of the total $B\rightarrow X_s\gamma\gamma$ sample. Since the
inclusive branching fraction of charmless \emph{B} events is roughly
estimated to be about 1\%, they are likely to provide the most
important background. However it is useful to note that a very
important characteristic of $b\rightarrow s\gamma\gamma$ is that the
photons carry off an appreciable fraction of the total energy, leaving
the mean energy of the \emph{s} quark to be around 1.7 GeV. Thus we
expect these $2\gamma$ events to be relatively clean with an average
multiplicity substantially less than in charmless $B$ events.  This
property should come in handy for separating the $2\gamma$
\emph{background} coming from the decay of $\pi^0$, $\eta$ and
$\eta^\prime$ from amongst the multibody charmless $B$ events. Another
remarkable feature of the prompt $b\rightarrow s\gamma\gamma$ signal
is that the photons have a large opening angle
($\langle\cos\theta_{\gamma\gamma}\rangle\simeq -0.7$), as is seen in
Fig.~\ref{dgammadcosth}. This is in sharp contrast to the $2\gamma$'s
from decays of the $\pi^0$, $\eta$ or $\eta^\prime$, which, as a rule,
should have a much smaller opening angle.

\section*{Acknowledgments}

We thank David Atwood, Hubert Simma and Ed Yao for discussions.  One
of us (G.R.) acknowledges the warm hospitality of Brookhaven National
Laboratory, during the first stage of this work.  This research was
supported in part by U.S. Department of Energy contract DE-AC-76CH0016
(BNL).


\begin{thebibliography}{99}
\bibitem{bsgg2} Work on QCD corrections, by the authors, is in
progress and will be published elsewhere.
\end{thebibliography}
\end{document}